\documentclass[12pt]{iopart}
\usepackage{graphicx}
\usepackage[T1]{fontenc}

\begin{document}
\title{Do the right thing}
\author{M.F. Laguna}\address{Centro At\'omico Bariloche, 8400 S.C. de Bariloche, Argentina}
\ead{lagunaf@cab.cnea.gov.ar}
\author{G. Abramson}\address{Centro At\'omico Bariloche and Instituto Balseiro, 8400 S.C. de Bariloche, Argentina}
\ead{abramson@cab.cnea.gov.ar}
\author{S. Risau-Gusman}\address{Centro At\'omico Bariloche, 8400 S.C. de Bariloche, Argentina}
\ead{srisau@cab.cnea.gov.ar}
\author{J. R. Iglesias}\address{Instituto de F\'{\i}sica and Faculdade de Ci\^encias Econ\'omicas, UFRGS, Caixa Postal 15051, 91501-970 Porto Alegre, RS, Brazil}
\ead{iglesias@if.ufrgs.br}

\begin{abstract}
We study a model of opinion formation where the opinions in conflict are not equivalent. This is the case when the subject of the decision is to respect a norm or a law. In such scenarios, one of the possible behaviors is to abide by the norm and the other to ignore it. The evolution of the dynamics is implemented through an imitation mechanism, in which agents can change their opinions based on the opinions of a set of partners and their own state. We determine, for different social situations, the minimum percentage of supporters of the law necessary to arrive to a state of consensus of law-abiders.
\end{abstract}

\pacs{89.65-s, 89.65.Ef}
\submitto{JSTAT}
\maketitle


\section{Introduction}
Elaborating on the seminal work of Axelrod on the dissemination of culture \cite{axelrod}, and the pioneering papers of Galam \cite{galam82,galam86}, a number of opinion formation models have been formulated in recent years. Their authors try to capture the fundamental processes that determine the emergency or not of consensus in a population (see an up-to-date review in Ref. \cite{castellano2009}).
Most of these studies assume that the agents must choose between equivalent options. However, in many social situations opinions may
be endowed with an intrinsic value derived from cultural, ethical or legal
foundations. Possible examples are the decision of using a cell phone while
driving a car, of stopping at the red light, of not driving after drinking, or no double park. One could also consider dietary, clothing and ritual prescriptions in many cultures. Even rules of etiquette that are not mandatory, in many cases make social coexistence easier. In all the preceding examples the choices are not equivalent, since there exist legal or customary values associated with them. In such a context, options or
opinions should be associated to a scale of values.
The decision to choose one or the other is greatly influenced by the agents'
background (education, respect of cultural values), their idiosyncrasy, the pressure of the social environment and of the authorities; it cannot be ruled out, also, that an agent could act on the spur of the moment. Eventually, not abiding by the law can result in some kind of penalty, such as a fine or even prison \cite{crimeandpunishment}.

In the theory of social impact, the influence between individuals is analyzed from a psychological point of view \cite{lantane1981}. In accordance with this view, the influence of a social group on an agent depends on the number of individuals in the group, on their persuasion power, and on the distance from the subject (either spatial or in the abstract sense of personal relationships). The interplay of both persuasion and authority in a hierarchical system was studied in \cite{authority}.
As we said, a fundamental contribution to this subject comes from the model introduced by Axelrod \cite{axelrod,axelrod1007}, which includes two fundamental mechanisms: social influence (tendency to became more similar with the interactions) and tendency of similar persons to attract each other and interact more frequently. The model predicts not only a convergence to a single culture but also persistence of diversity in some particular cases.
Finally, the resistance of a society to the convergence of its members' opinions has been considered in \cite{laguna2004}, and a mean field approach of the ``stubborn'' society has been studied in \cite{porfiri2007}. In both cases a homogeneously stubborn population was considered, while individual variation of obstinacy is a more reasonable approach, and it is the one we follow here.

In this paper we study a simplified instance of such a problem. We consider a system where the agents may adopt one of two options with different
values. We take into account the fact that social influence can play a major role in the choice of a given opinion. As a model of such influence, we propose an imitation behavior, where agents tend to imitate the behavior of the others, which may be the right or the wrong one. This is the approach followed in other recent studies of the formation and propagation of opinion, such as the herding model found in \cite{schwammle2007} and the voting model of \cite{moreira2006}. As a result of this influence, agents can change their opinion. At variance with models where the opinions are equivalent, the choice will here depend not only on the agent's obstinacy and number of social contacts, but also on the value of their opinions. 

We will show that in many cases, it is possible to induce a major change in the opinion of the population by setting a fraction of \emph{polite} or \emph{educated} agents (those supporting the ``right'' option, acting like the ``seeds'' in \cite{moreira2006}). Their influence will be able to generate a change in the society in a reasonable period of time. The fraction of educated agents necessary to induce a global change in opinion depends on the relative value of the opinions, on the average adamancy and on the connectivity of the network.

In the next section we describe the model while in section~\ref{res} the results will be presented for different scenarios. Discussion of the results and conclusions follow in section~\ref{conclu}.


\section{Model of Opinion Dynamics}
\label{model}

Let us consider a population of $N$ interacting agents, each of them having an
{\it opinion} attribute and a tendency to conserve this opinion, that we call {\it adamancy}. We assume that the opinion of each agent has a ``value'' denoted by $x$, being $x \in \{x_1,x_2\}$, where $x_2>x_1$, implying that opinion 2 is more valuable than opinion 1. For example opinion $2$ can be identified with abiding by a given rule, while opinion $1$ means a tendency to ignore
it. Individuals are also characterized by an adamancy factor $a$, which they keep for the whole time evolution, measuring their resistance to change their current opinion\footnote{If $a<1$, the same parameter could serve as the opposite of stubbornness, since one's own opinion weighs less than that of others.}. We have considered both the case of an homogeneous adamancy and of a randomly distributed one. Finally, agents may change their opinions as a result of their mutual
interactions.

The dynamics of the system is as follows. At each time step an agent $i$ is
picked up at random. This agent will be prompted to change its opinion because
of the influence of a group of $k$ individuals. These are chosen at random at each interaction event in the current well-mixed implementation of the model. After the selection of this group of
influence, a poll is performed:
\begin{equation}
\label{poll}
\mbox{If: } a_i x_i +n_i x_i-(k-n_i)x'_i 
\left\{
\begin{array}{ll}
\geq 0 & \mbox{then agent $i$ keeps opinion $x_i$},\\ 
<0     & \mbox{then agent $i$ changes to opinion $x'_i$,}
\end{array}
\right.
\end{equation}
where $n_i$ is the number of agents (belonging to the influence
group of size $k$) sharing the same opinion as agent $i$, and $x'_i$ is the
alternative opinion sustained by the rest of the agents: $k-n_i$.

It is possible to see that there is a competition between the two opinion values. On
the one side we have the agents with the same opinion as agent $i$, and on the
opposite side the agents with the alternative opinion. In addition, we see that agent $i$
weighs his own opinion (and only this one) with his adamancy $a_i$. As an example, let us suppose that $x_1=1$ and $x_2=2$. Let us say 
that agent $i$ has the lower opinion, i.e., $x_{i} = 1$. If we set $k=4$ (four agents in the group of influence), and we consider that two of them share the lower opinion, $x=1$, whereas
the other two have the higher opinion value $x=2$, then opinion 1 will sum $O_{1} = a_i \times 1 + 2 \times 1 = a_i+2$, whereas opinion 2 will sum $O_{2} = 2 \times
2 = 4$. If $a_i<2$ then $O_{1}<O_{2}$, opinion 2 will be the winner of the
poll and agent $i$ will change his opinion from $1$ to $2$. Thus, in this case,
agent $i$ will adopt the higher valued opinion.

This description emphasizes the weight of the agent's own opinion, with a central role played by his adamancy or stubbornness. This is an important point of view of human psychology and behavior, and for this reason we chose to write Eq.~(\ref{poll}) in this way. Observe, however, that one can alternatively scale the local parameters with the size of the influence group $k_i$. Defining $w_i=a_i/k_i$ as the weight of the agent's own opinion in the poll (his adamancy ``diluted'' among his group of influence), and dividing Eq.~(\ref{poll}) by $k_i$ we can write:
\begin{equation}
\label{poll2}
\mbox{If: } w_i x_i +s_i x_i-(1-s_i)x'_i 
\left\{
\begin{array}{ll}
\geq 0 & \mbox{then agent $i$ keeps opinion $x_i$},\\ 
<0     & \mbox{then agent $i$ changes to opinion $x'_i$,}
\end{array}
\right.
\end{equation}
where $s_i=n_i/k_i$ is the fraction of agent's $i$ partners supporting the same opinion as his. The number of parameters in the dynamics described by Eq.~(\ref{poll2}) is the same as in Eq.~(\ref{poll}), and a change of perspective is achieved: the competition between the weighted opinions scales with the size of the group of influence, and in this sense it becomes uniformly defined in the population.

The process defined by Eqs.~(\ref{poll}) or (\ref{poll2}) is then iterated by selecting agents at random. The final state will depend on the level and distribution of adamancy $a$ of the population, on the fraction $f_h$ of agents that have initially the higher opinion, and also on the distribution of $k$. 

The dynamics just described has only two attractors in a strict sense: consensus of opinion 1 or consensus of opinion 2. Nevertheless, the time needed to reach such stable states can be extremely long. For intermediate times, as we will show in the next section, the system seems to have metastable states in which the two opinions coexist in the population.

There is a note to be made about the relative value of the opinions. In the example we fixed the value of the higher (desirable) opinion as double that of the other, and this is also the case that we will analyze in most of our results below. This choice is arbitrary and could be different in different real situations. In general, laws and norms are created to rule unresolved conflicts. For example, the priority on a crossing to avoid accidents, the prohibition to use cell phones while driving or of smoking in closed areas belong to this class. It is clear that one of the opinions
has a higher value, but it is not so clear how to quantify it. We adopt for
the moment the values $x_1=1$ and $x_2=2$, and in the final section we
discuss the effect of choosing different numerical values for the opinions.
Anticipating on the results, different cases were observed: for some values of
$a$ and $f_h$ the system arrives to a consensus, where all agents share the same opinion. When the adamancy is high enough, a frozen state is achieved in which the two opinions coexist in the population.

In the following section we present our results for different distributions of the
adamancy $a$ and the average connectivity $\langle k\rangle$.

\section{Results}
\label{res}

We have analyzed several cases, from the simplest ones to more ``realistic'' situations. We start with a simple situation consisting of an homogeneous adamancy and a fixed number of agents in the group of influence. Then, we consider a random distribution for the adamancy, keeping the number of connections constant and the same for all agents. Finally we describe the case where the number of connections is random, but the adamancy is the same for all the agents. In our numerical simulations the value of the opinions will be $1$ and $2$, while a more general case with a continuous of opinion values is exhaustively analyzed in the last section. The system size was fixed at $N=1000$ agents since no significant dependence with size was observed (at variance with the models analyzed in \cite{tessone}). The \emph{time} necessary to reach the final states \emph{does} depend on the system size, naturally; but neither the phase diagrams nor the temporal evolution of the populations of each state do. We will discuss the final times of evolution more in depth below, at the end of  Section \ref{res-homo}.

\subsection{Homogeneous adamancy and connectivity}
\label{res-homo}

Consider a situation where each agent interacts with the same number of partners, chosen at random. The simplest case is a pairwise interaction, i.e. each agent interacts with only another one. If both interacting agents share the same opinion, there are no changes. On the other hand, we have that either:
\begin{itemize}
\item Agent $i$ supports opinion 1 ($x_i=1$). If $a_i < 2$, agent $i$ will be convinced by $j$ to change his opinion from 1 to 2.
\end{itemize}
or:
\begin{itemize}
\item Agent $i$ supports opinion 2 ($x_i=2$). If $a_i < 0.5$, agent $i$ will be convinced by $j$ to change his opinion from 2 to 1.
\end{itemize}

\begin{figure*}[t]
        \begin{center}
        \includegraphics[width=\columnwidth]{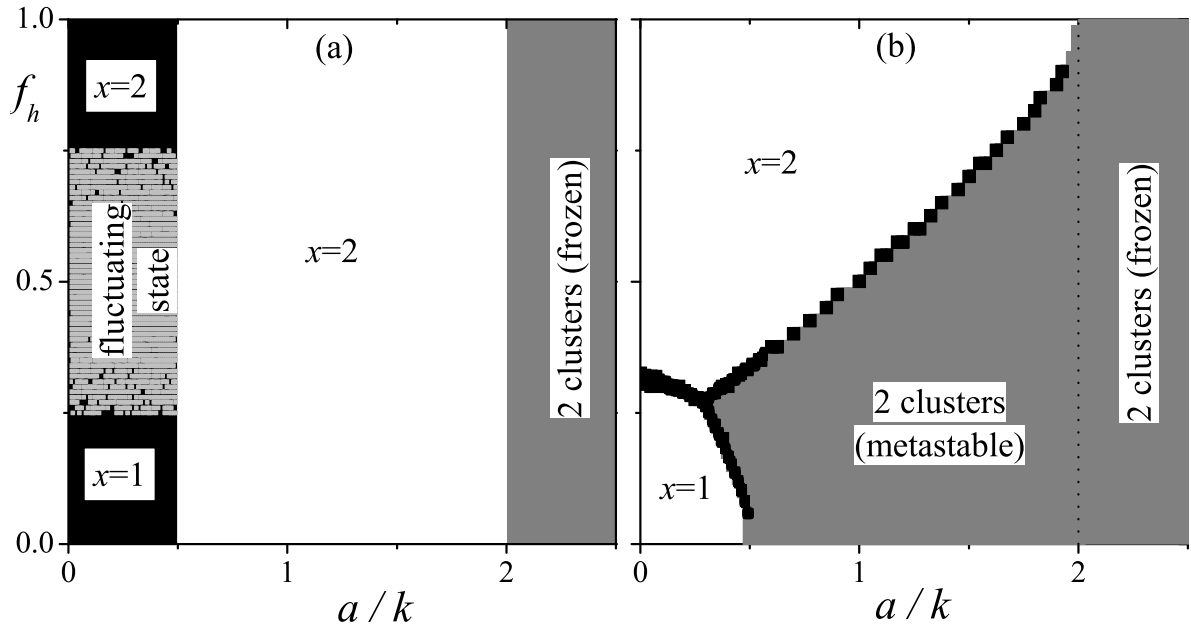}
        \end{center}
        \caption{Phase diagrams for $k=1$ (a) and $k=100$ (b). Regions colored grey correspond to states of coexistence of opinions. White corresponds to consensus of either opinion, as shown. Black indicates a transition region. Finally, light gray in panel (a) indicates a fluctuating state.
All simulations consist of 100 realizations per point in the phase diagrams. Note that the horizontal axis represents the weight of the agent opinion $w\equiv a/k$.}
\label{fig:Kconst}
\end{figure*}

The consequences of these two conditions can be seen in Fig. \ref{fig:Kconst}(a) as vertical lines. The phase diagram shows the final state of the system in a space defined by the scaled adamancy $w=a/k$ and the fraction $f_h$ of initially supporters of the higher valued opinion $x=2$. 
The region with adamancy lower than 0.5 corresponds to the most voluble populations. In this region, the agents convince each other at random. Therefore, the fraction of agents supporting each opinion fluctuates stochastically until the population collapses to a state of consensus. On average, the time needed for this collapse and the value of the consensus opinion depend on the initial state. For example, an initial condition with a large fraction of individuals holding one of the opinions will collapse quickly to a state of consensus of the same opinion, with high probability (black regions in Fig. \ref{fig:Kconst}(a)). When the initial fractions of agents holding each opinion are not very different (light grey region in Fig. \ref{fig:Kconst}(a)), the time to reach consensus can be much longer than the one used in the simulations (fixed in 100 interactions per agent, which is a reasonable assumption in the context of social interactions). When the scaled adamancy has intermediate values ($0.5<w<2$) the final state is a consensus of opinion $x=2$ (white region in Fig.~\ref{fig:Kconst}). Finally, for very stubborn populations there is coexistence of opinions for all values of $f_h$, reflecting the fact that no interaction ends up in a change of opinion. This is a ``frozen'' stable state, since for high values of the stubbornness no agents modify their opinion. 

On the other hand, if we increase the number of interacting agents (chosen at random in each interaction step) the situation is different. The corresponding phase diagram is shown in  Fig.~\ref{fig:Kconst}(b), for $k=100$ (which is close to a mean field scenario in our system of $N=1000$ agents).  It is possible to see that the higher opinion is always the winner if the number of educated people is greater than a critical value that depends on the adamancy of the population. This critical $f_h$ (painted black in the figure) grows with adamancy---as expected---for $w>0.5$. Observe that there is a minimum of the critical $f_h$ around $w \approx 0.3$, where a fraction of educated agents ($30\%$) is enough to convert the whole population to opinion 2. We remark that, as the abscissa is $w=a/k$, this condition is attained for any reasonable values of the adamancy. Below the critical $f_h$ we find a region of two-clusters states, which is metastable with a slow dynamics. For $w>2$ all the two-clusters states become stable, frozen at the initial condition---as in the case $k=1$. Finally, a region of prevailing opinion 1 occupies the lower-left corner of low adamancy and little initially educated agents. In this region, a larger value of $f_h$ is necessary to impose opinion 2 for lower values of adamancy. 

\begin{figure*}[tb]
        \begin{center}
        \includegraphics[width=\columnwidth]{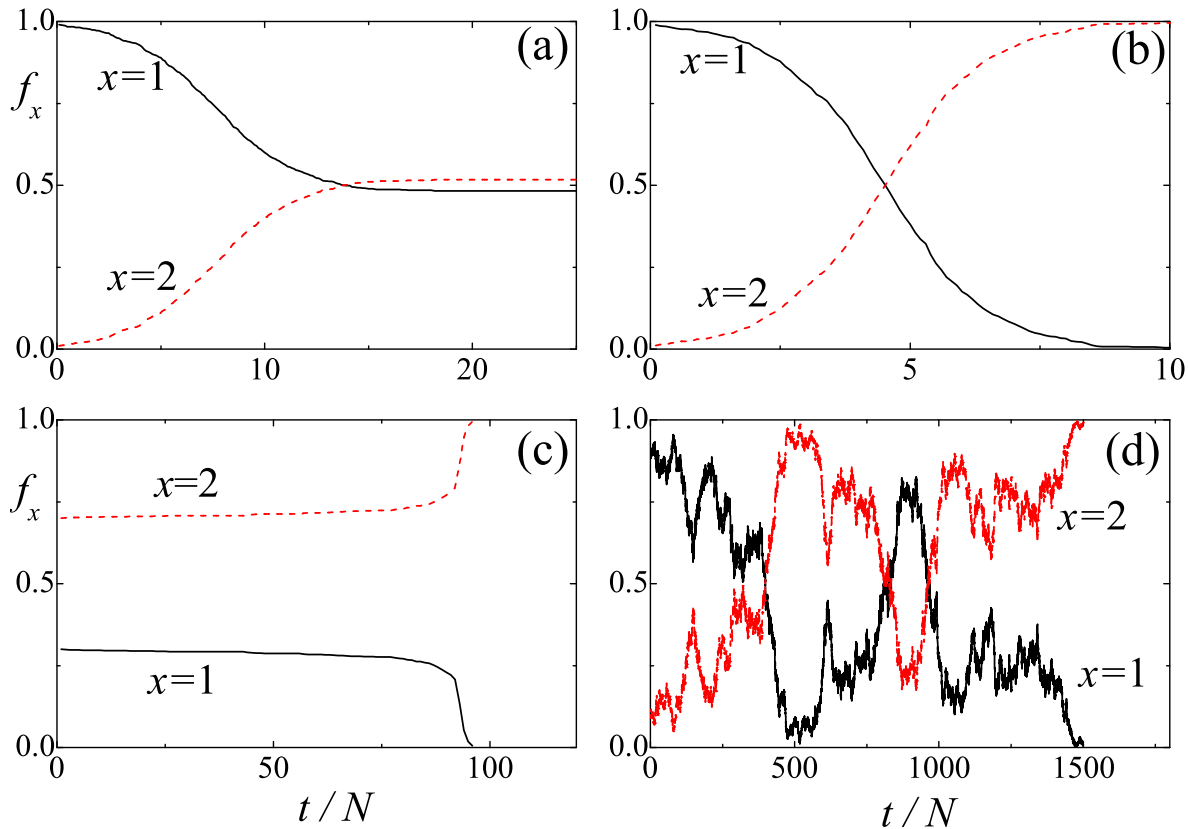}
        \end{center}
        \caption{Evolution of the system state, corresponding to four points in parameter space: (a) $f_h=0.01$, $k=1$, $\langle a\rangle=2$, $\sigma_a=0.2$ (this is one of the heterogeneous systems analyzed below in Section~\ref{hetero}), (b) $f_h=0.01$, $a=1.8$, $k=1$, (c) $f_h=0.7$, $a=150$, $k=100$, (d) $f_h=0.1$, $a=0.3$, $k=1$.}
\label{fig:evol1}
\end{figure*} 

We have observed that intermediate values of $k$ show a behavior qualitatively similar to the one shown for $k=100$ in Fig.~\ref{fig:Kconst}(b). Only the case $k=1$ stands out as qualitatively different, as described. Below, in Fig.~\ref{fig:kvar}, we will show some results for one of the intermediate cases, $k=10$.

The results just discussed correspond to final states of the dynamics accessible during the numerical simulation. We have observed that many situations present very long transient times, and since the model pretends to mimic a social system, it could be helpful to address the question of the time needed to achieve the final state. 
In a real society the process to reach a state of consensus cannot take an unreasonable large number of interactions. This is the reason why we choose a finite number of interactions (100 per agent, in average) to construct the phase diagrams, even though in some regions the stable states are not the ones observed in our simulations. As an extreme, we note that both consensus are stable almost anywhere, since they cannot be invaded by a minimal perturbation consisting of a single agent of the opposing opinion. 
To better understand the dynamic behavior of our system, observe the transients shown in Fig.~\ref{fig:evol1}. On Fig. \ref{fig:evol1}(a) we see the system rapidly evolving into a coexisting state of opinions 1 and 2. Fig.~\ref{fig:evol1}(b) shows a similarly fast evolution towards a consensus of opinion 2. Figure~\ref{fig:evol1}(c) shows a much longer transient state, corresponding to the transition between the region of consensus of opinion 2 and the metastable region of coexistence (the thin black region in Fig.~\ref{fig:Kconst}(b)). Finally, Fig.~\ref{fig:evol1}(d) shows the slow evolution of the low adamancy region of Fig. \ref{fig:Kconst}(a): from a practical point of view, this society is living in a (fluctuating) state of coexisting opinions.

\begin{figure*}
        \begin{center}
        \includegraphics[width=\columnwidth]{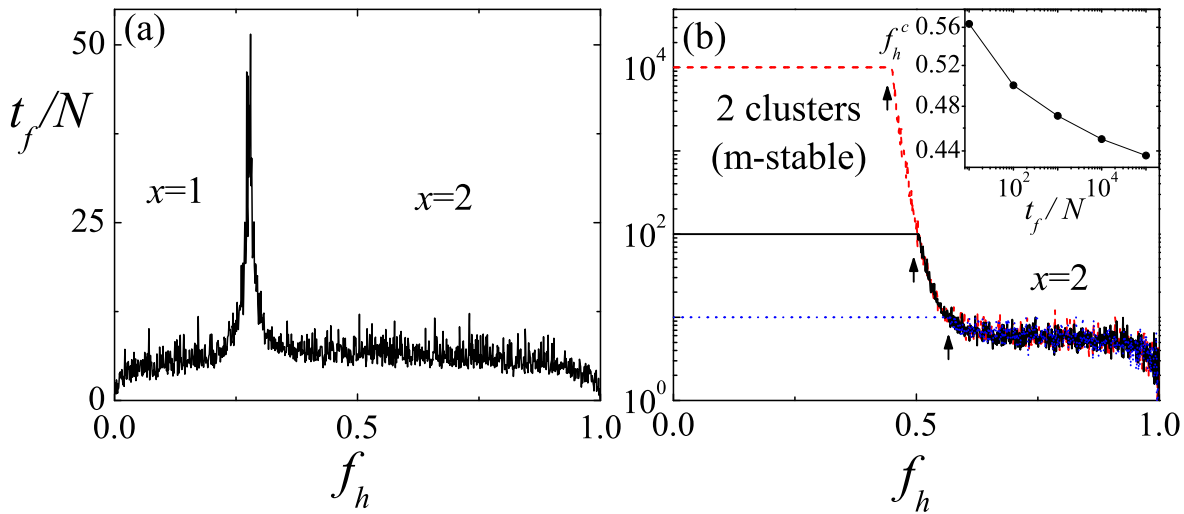}
        \end{center}
        \caption{Necessary time to reach the stationary state, as a function of the fraction of agents initially endowed with opinion $x=2$. We plot the scaled final time ($t_f/N$, number of interactions per agent) for $k=100$ and (a) $a=100$; (b) $a=25$, with the arrows indicating the value of $f_h$ at which the transition occurs. Inset: Critical $f_h$ as a function of $t_f/N$ (note that the decay is extremely slow, in fact slower than algebraic, and that the scale is double logarithmic in spite of the fact that the small range in the vertical axis hinders its appreciation).}
\label{fig:transient}
\end{figure*}

Observe also Fig.~\ref{fig:transient}, which shows the time necessary to reach the final state, as a function of $f_h$, for a given value of adamancy. Panel (a) corresponds to $k=100$ and $a=0.25k$, while panel (b) shows the case $k=100$, $a=k$, both corresponding to vertical cuts in Fig.~\ref{fig:Kconst}(b). The first one shows a peak corresponding to the transition between the two regions of consensus. The second one shows the transition between the region of two clusters and the region of ``right'' ($x=2$) consensus. One sees that well within the region of consensus the system evolves rapidly to reach the final state. Going from higher to lower values of $f_h$ and close to the transition line, the transient time grows abruptly (corresponding to the slow evolution seen in Fig.~\ref{fig:evol1}(c)). Finally, for $f_h<0.5$, the maximum simulation time is reached without abandoning the two-clusters state (but reaching a stationary state in a short time, as seen in Fig.~\ref{fig:evol1}(a)). 
In the previous figures we used the criterion of 100 interactions per agent as maximum simulation time. This case is shown as a full line in Fig.~\ref{fig:transient}(b). The two other lines correspond to two different choices of this criterion: the dotted line (smaller times) corresponds to 10 interactions per agent, while the dashed one (much longer times, note the logarithmic scale) to $10^4$ interactions per agent. Next to these lines, the arrows indicate the value of $f_h$ at which the transition occurs. This value is plotted as a function of the final time in the inset, decaying slower than a power law.

\subsection{Heterogeneous systems}
\label{hetero} 

A homogeneous adamancy in the system, namely that all agents are equally stubborn in preserving their opinions, is surely a very strong assumption for a human population. In this context, we have relaxed the condition of homogeneity by allowing a distribution of adamancy in the population. A random value of the adamancy is drawn initially from a square distribution for each agent. We have used the mean value of this distribution as one of the parameters for the phase diagrams, equivalent to the parameter $a$ used in Fig.~\ref{fig:Kconst}. In these systems we have observed that the phase diagrams are substantially unchanged with respect to those of homogeneous adamancy, at least when the width of the adamancy distribution is within reasonable limits. The only difference to be found is a narrow coexistence phase that appears in the transition lines between phases. One such case is exemplified by its temporal evolution in Fig.~\ref{fig:evol1}(a), where a system with a distribution of adamancy of width $\sigma_a=0.2$ evolves to a mixed state of both opinions.

\begin{figure*}
        \begin{center}
        \includegraphics[width=\columnwidth]{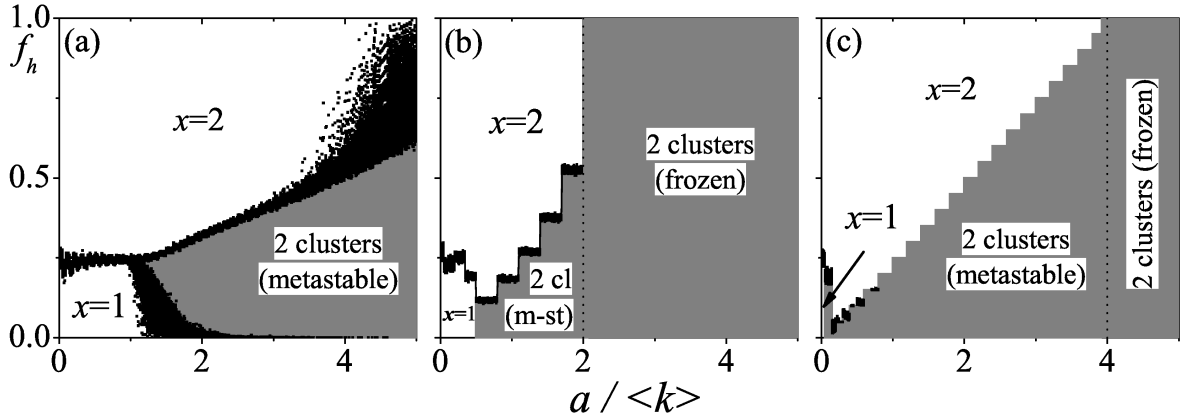}
        \end{center}
        \caption{Phase diagrams for different distributions of the connectivity, all with $\langle k\rangle=10$. (a) exponential distribution; (b) delta distribution (homogeneous case) and (c) square (width 20). Colored as Fig.~\ref{fig:Kconst}. Observe that the horizontal axis represents a weight ---an ``effective'' one in this case--- of the agent's opinion, as in Fig.~\ref{fig:Kconst}.}
        \label{fig:kvar}
\end{figure*}

On the other hand, and with the same spirit, we have explored systems that have an heterogeneous connectivity and homogeneous adamancy. In this case, we chose to draw a random value of the number of partners at each interaction step. Two distributions of this variable were considered: an exponentially decaying one and a square one (this last defined by $p(k)=(2\langle k\rangle )^{-1}$, $k\in [0,2\langle k\rangle ]$). The results also show little variation with respect to that of the completely homogeneous system (delta distribution at the value $\langle k\rangle$), as one can see in Fig. \ref{fig:kvar}. These plots show the two mentioned $k$-distributed cases with $\langle k\rangle =10$, next to the homogeneous one for $k=10$ (this is the intermediate case between $k=1$ and 100, referred to above). The general picture of three regions persists: a consensus of opinion 2 above a critical $f_h$, a small region of opinion 1 cornered in the region of low adamancy and few initially educated agents, and a broad region of coexistence in the complement. We have also explored power-law distributions of $k$, and observed that the phase space remains mostly unchanged. Indeed, an algebraic distribution of $k$ appropriately chosen to have $\langle k\rangle = 10$ produces a phase space nearly identical to the one corresponding to the exponential on Fig.~\ref{fig:kvar}.

The exponential and the square distributions provide contrasting effects in the dynamics of opinion imitation. Observe that the exponential one has an excess of \emph{small} connectivities with respect to the square one. On the other hand, the square distribution has an excess of \emph{high} connectivities with respect to the exponential. When there is an excess of agents with small connectivity (exponential distribution of $k$), it is difficult for those of them supporting opinion 1 to be gained by a pool of opinion 2 partners: the weight of the few partners may just not be enough. As a result of this, the region of consensus of opinion 1 is wider. On the other hand, when there is an excess of large connectivities, the random interactions are able to eventually provide enough partners with opinion 2. This makes the region of consensus $x=1$ shrink.

\section{Discussion and conclusions}
\label{conclu}

As mentioned in the introduction, the numerical value of the opinions is a free parameter of the model. A simple analysis can show that these values have a direct consequence on  the threshold of $f_h$, above which a consensus of the high valued opinion emerges. For the sake of simplicity, let us consider $x_1 = 1$ and $x_2>1$. It is easy to obtain an evolution equation for $f_h$, the fraction of agents holding opinion 2:
\begin{equation}
\label{eq:evol}
\dot{f_h} = (1-f_h) \sum_{i=\lceil i_k \rceil}^k {k \choose i} f_h^i (1-f_h)^{k-i}  - f_h \sum_{i=\lceil i_k x_2\rceil}^k {k \choose i} f_h^{k-i} (1-f_h)^i ,
\end{equation}
where $i_k=(a+k)/(1+x_2)$ and $\lceil i_k \rceil$ is the smallest integer larger than $i_k$, which gives the smallest number of  agents holding opinion $2$ that are necessary to convince an agent holding opinion $1$. $\lceil i_k x_2\rceil$ gives the smallest number of agents holding opinion $1$ that are necessary to convince an agent holding opinion $2$. Equation (\ref{eq:evol}) can be recast as:
\begin{equation}
\dot{f_h} = f_h (1-f_h)^k \left[ \sum_{i=\lceil i_k \rceil}^k {k \choose i} f^{i-1} - \sum_{i=\lceil i_k x_2\rceil}^k {k \choose i} f^{k-i} \right] \equiv f_h (1-f_h)^k P_k(f),
\end{equation}
where $f \equiv f_h/(1-f_h)$. Note first that, if $i_k>k$ (that is, if $a>k x_2$) we get $P_k(f) \equiv 0$, and there is no evolution from the initial state (as found in the simulations). If $\lceil i_k x_2\rceil > k$ (that is, if $k/x_2<a<k x_2$) we get $P_k(f)>0$ for all $f$ and therefore the system evolves to a consensus of opinion $2$ for all values of $f$. For $a<k/x_2$ the different cases for each value of $k$ must be analyzed separately.

The simplest nontrivial case is the one with $k=2$. To convince an agent holding opinion $2$, it is evident that we need $2$ agents holding opinion $1$, and therefore $\lceil i_2 x_2\rceil =2$. We have two subcases now: $x_2 \geq 2$ and $x_2<2$. In the first subcase we obtain $\lceil i_2 \rceil =1$, and therefore $P_k(f)=f+1>1$. In other words, the equilibrium state is consensus of opinion $2$ for all values of the adamancy. On the other hand, if $x_2<2$ we have that if $a>x_2-1$, then $\lceil i_2 \rceil =1$ (which implies consensus of opinion $2$, as in the previous subcase), and $a<x_2-1$ then $\lceil i_2 \rceil =2$. In this last case we have $P_k(f)=f-1$. Thus, there is an unstable equilibrium if $f=1$, i.e. $f_h=1/2$. If the initial value of $f_h$ is smaller than $1/2$ then system converges to a consensus of opinion $1$; otherwise, it converges to a consensus of opinion $2$.

\begin{figure*}
        \begin{center}
        \includegraphics[width=\columnwidth]{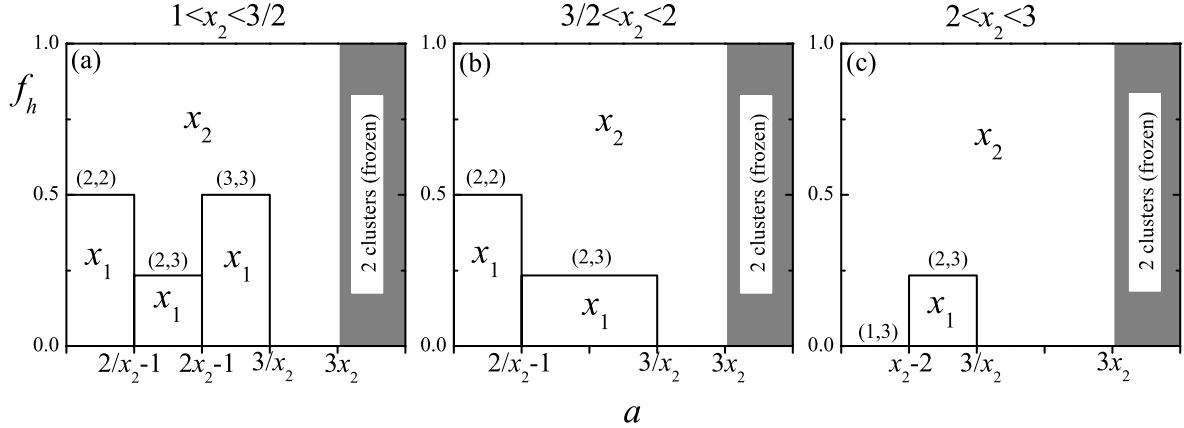}
        \end{center}
        \caption{Analytical phase diagrams for connectivity $k=3$, $x_1=1$ and a continuous value of $x_2$ (with $x_1<x_2$). The labels on top of each panel indicate the corresponding range of $x_2$. Two stationary states are found for $a<kx_2$: consensus of opinion $x_2$ and consensus of opinion $x_1$. For higher values of $a$ a frozen state is obtained. In parentheses, the first (second) number indicates the minimum number of agents with opinion $x_2$ ($x_1$) needed to convince an agent with opinion $x_1$ ($x_2$).}
        \label{fig:analitica}
\end{figure*}

For $k>2$ the analysis sketched above becomes increasingly more involved. As an example, Figure~\ref{fig:analitica} shows the three cases with connectivity $k=3$ corresponding to all the possible relations between the two opinion values. Some regions in these diagrams are simple to understand: as before, if $a>k x_2$ then $P_k(f)=0$ and the initial state is frozen. In the region $a<k/x_2$ there are two possible outcomes: either consensus of $x_1$ or of $x_2$, separated at a critical value of $f_h$ that depends on the adamancy $a$. The possible situations are represented as columns in the figure, each one identified by a pair of numbers in parentheses. For example, in the center panel, the left hand column reflects the fact that if adamancy is $0<a<2/x_2-1$, the system is attracted by opinion $x_2$ if more than half the population initially supports it, and by opinion $x_1$ otherwise. This means that the transition occurs at the value $f_h=0.5$. In parentheses,  the first number indicates the minimum number of agents with $x_2$ needed for an agent with $x_1$ to be convinced, and the second number shows the same, interchanging 1 and 2. Since these numbers are discrete, a structure of columns appears even though adamancy is a continuous parameter. Observe that this structure is also seen in the simulations, for example in Fig.~\ref{fig:kvar}.

In almost all the phase diagrams an ``island'' of $x=1$ consensus is observed in the lower left corner of parameter space (Figs.~\ref{fig:Kconst}, \ref{fig:kvar} and  \ref{fig:analitica}, with the exception of the case shown in Fig.~\ref{fig:analitica}(c), corresponding to very high values of $x_2$). Consider for example a case with adamancy $a=0$. In this case, an agent with opinion $x=1$ will preserve it if twice as many agents with opinion 1 exist in his group of influence, for each agent supporting opinion 2. As a consequence, in average the transition will happen for $f_h=1/3$, as observed in Fig.~\ref{fig:Kconst}(b). The transition happens at a different value of $f_h$ if the relative value of the opinions is different. For example, if $x_2=3$ the threshold is $f_h=0.25$, if $x_2=4$, $f_h=0.2$, and so on. Then, the minimum fraction of educated agents needed to reach consensus of the ``right'' opinion depends on the relative value of the opinions. That is, a high valued opinion will need a relatively small number of educated people to become the ruling one. On the other hand, and as expected, the stubbornness of the population constitute a negative factor when trying to spread the ``right'' opinion. 

One can see here what at first looks like a paradoxical effect. Observe for example Fig.~\ref{fig:kvar}, where three systems with $\langle k\rangle=10$ are shown side by side. we see that the worst situation---meaning that with the least area occupied by opinion 2 in parameter space---is the case in which the distribution of adamancy is a $\delta$-function (panel (b)). In this case, indeed, it is impossible to change the mind of the population if the adamancy is greater than $2\langle k\rangle$. On the other hand, in panel (a) we see that an exponentially distributed adamancy has a significantly larger region of consensus of opinion 2, up to values of the adamancy $a\sim 4\langle k\rangle$. The exponential distribution has its maximum at $a=0$, thus providing most of the times \emph{less} contacts to the interacting agents. However, its tail provides \emph{more} contacts than the average. More contacts make it easier for supporters of opinion 1 to change their minds by imitation. This process acts effectively as a source of  supporters of opinion 2, improving in the end the behavior of the whole population. Finally, we also observe that the region of consensus of opinion 1 is smaller for a delta distribution of adamancy than for an exponential, and even smaller for the uniform distribution.

In any case, the number of acquaintances of an agent in a social environment is not easy to establish. In some situations one could assume a power-law distribution as in small worlds. Yet, in other scenarios of social interest these distributions would not be realistic: consider for example people driving a car in an urban environment, where the average number of other drivers would be the same for all the agents. If an exponential distribution of the number of contacts seems to be favorable to disseminate the ``right'' behavior, it can also stabilize the ``bad'' behavior if the initial fraction of educated people is small. On the other hand a delta distribution, that seems to reproduce better the situation of drivers in a city, exhibits a relatively reduced region where the society converges to the low valued opinion, but can easily lead to a two clusters state, as it is observed in many real situations.

In all cases the role of education is crucial: education that can act both through persuasion of a ``core'' of the population as well as by the punishment of the ``wrong'' behavior. But both persuasion or punishment may also have a negative effect: to generate a more adamant behavior by imposing too high a pressure on agents not abiding with the right behavior.

We are aware of the simplicity of the present model, but we think that it provides a nice picture of the problem of imposing a given behavior in the society. Further steps will be taken in following studies, in particular along the lines presented in~\cite{crimeandpunishment}, including the effects of punishment as an additional ingredient in the dynamics. The role of noise, which is known to play an important role in the formation of consensus (see for example~\cite{schwammle2007}) will also be subject to investigation.

\ack{
The authors acknowledge support from projects CAPES-MINCyT \#151/08 - 017/07 (Brazil-Argentina) and Project CNPq/Prosul 490440/2007. G.A. also acknowledges partial support from ANPCYT (PICT 04/943) and CONICET (PIP112-200801-00076) and J.R.I. to CNPq. We are also grateful with S. Gon\c calves for fruitful discussions.}

\end{document}